\expandafter\def\csname ver@fixltx2e.sty\endcsname{}
\PassOptionsToPackage{pdfpagelabels=false}{hyperref}

\documentclass[useAMS,usenatbib]{mnras}

\usepackage{amsfonts,amsmath,amssymb,mathrsfs}
\usepackage{graphicx,epsf,epsfig}
\usepackage{bm}
\usepackage{longtable}
\usepackage[usenames,dvipsnames]{xcolor}
\usepackage{multirow}
\usepackage{graphicx}

\usepackage[utf8]{inputenc}
\usepackage{hyperref}
\usepackage[hyphenbreaks]{breakurl}

\usepackage{times}
\usepackage{color}
\usepackage{breakurl}
\usepackage{color}
\usepackage{mathrsfs}

\hypersetup{draft}

\title[Time evolution of rotating and magnetized white dwarf stars]{Time evolution of rotating and magnetized white dwarf stars}

\author[L.~Becerra, K.~Boshkayev, J.~A.~Rueda and R.~Ruffini]{L.~Becerra,$^{1,2}$\thanks{laura.becerra@icranet.org} K.~Boshkayev,$^{3,4}$\thanks{kuantay.boshkayev@nu.edu.kz, kuantay@mail.ru} J.~A.~Rueda$^{1,2,5}$\thanks{jorge.rueda@icra.it}  R.~Ruffini$^{1,2,5}$\thanks{ruffini@icra.it}\\
$^1$ICRANet, Piazza della Repubblica 10, I--65122 Pescara, Italy\\
$^2$Dipartimento di Fisica and ICRA, Sapienza Universit\`a di Roma, P.le Aldo Moro 5, I--00185 Rome, Italy\\
$^3$NNLOT, al-Farabi Kazakh National University, Al-Farabi av. 71, 050040 Almaty, Kazakhstan\\
$^4$Department of Physics, Nazarbayev University, Kabanbay Batyr 53, 010000 Astana, Kazakhstan\\
$^5$ICRANet-Rio, CBPF, Rua Dr. Xavier Sigaud 150, Rio de Janeiro, RJ, 22290--180, Brazil}

\begin{document}

\date{\today}
\maketitle

\begin{abstract}

We investigate the evolution of isolated, zero and finite temperature, massive, uniformly rotating and highly magnetized white dwarf stars under angular momentum loss driven by magnetic dipole braking. We consider the structure and thermal evolution of white dwarf isothermal cores taking also into account the nuclear burning and neutrino emission processes. We estimate the white dwarf lifetime before it reaches the condition either for a type Ia supernova explosion or for the gravitational collapse to a neutron star. We study white dwarfs with surface magnetic fields from $10^6$ to $10^{9}$~G and masses from $1.39$ to $1.46~M_\odot$ and analyze the behavior of the white dwarf parameters such as moment of inertia, angular momentum, central temperature and magnetic field intensity as a function of lifetime.  The magnetic field is involved only to slow down white dwarfs, without affecting their equation of state and structure. In addition, we compute the characteristic time of nuclear reactions and dynamical time scale. The astrophysical consequences of the results are discussed.

\end{abstract}


\section{Introduction}\label{sec:1}

In this work we investigate the time evolution of massive, uniformly rotating, highly magnetized white dwarfs (WDs) when they lose angular momentum owing to magnetic dipole braking. We have different important reasons to perform such an investigation: 1) there is incontestable observational data on the existence of massive ($M\sim 1~M_\odot$) WDs with magnetic fields all the way to $10^{9}$~G (see \citep{2015MNRAS.446.4078K}, and references therein); 2) WDs can rotate with periods as short as $P\approx 0.5$~s \citep{2013ApJ...762..117B}; 3) they can be formed in double WD mergers \citep{2012ApJ...749...25G,2013ApJ...772L..24R,2018MNRAS.479L.113K}; 4) they have been invoked to explain type Ia supernovae within the \emph{double degenerate scenario} \citep{1984ApJ...277..355W,1984ApJS...54..335I}; 5) they constitute a viable model to explain soft gamma-repeaters and anomalous X-ray pulsars \citep{2012PASJ...64...56M,2013A&A...555A.151B,2013ApJ...772L..24R, 2015mgm..conf.2295B,2015NuPhA.937...17B}.

Thus, it is of astrophysical importance to determine qualitatively and quantitatively the evolution of micro and macro physical properties such as density, pressure, temperature, mass, radius, moment of inertia, and angular velocity/rotation period of massive, uniformly rotating, highly magnetized WDs. We constrain ourselves in this work to study the specific case when the WD is isolated and it is losing angular momentum via magnetic dipole braking. In addition, we explore WDs with surface magnetic fields from $10^6$ to $10^{9}$~G and masses from $1.39$ to $1.46~M_\odot$.

Depending on their mass, WDs can exhibit different timing properties by losing angular momentum: uniformly and differentially rotating super-Chandrasekhar WDs (hereafter SCWDs) spin-up, whereas sub-Chandrasekhar WDs spin-down. The possibility that a rotating star spin-up by angular momentum loss was first revealed by \citet{1990ApJ...357L..17S}, and later by \citet{2000ApJ...534..359G}. In both articles the WDs were studied within the Newtonian framework, though the effects of general relativity are crucial to determine the stability of massive, fast rotating WDs (see, e.g., \citep{shapirobook,2011PhRvD..84h4007R,2013ApJ...762..117B,2015mgm..conf.2468B,2016JPhCS.706e2016C,2018GReGr..50...38C,2018mgm..conf.4319C}). Besides the above timing features, we shall quantify the compression that the WD suffers while evolving via angular momentum loss which is relevant for some models of type Ia supernovae.

Specifically, we focus on the compression of isolated zero-temperature super-Chandrasekhar WDs and their isothermal cores by angular momentum loss based on our previous results \citep{2013ApJ...762..117B,2014IJMPKazNU,2014JKPS...65..855B,2017mgm}.
We shall estimate the lifetime ($\tau$) of zero temperature WDs via magnetic dipole braking taking into due account the effect of the time evolution of all relevant parameters of the WDs. In addition, for hot isothermal WD cores we compute the times of the nuclear burning and neutrino emission phenomena. For both cold and hot WDs, we consider two magnetic field evolution: (1) a constant magnetic field during the entire evolution and, (2) a varying magnetic field according to magnetic flux conservation \citep{2014IJMPKazNU,2014JKPS...65..855B,2017mgm}.

The influence of the extreme magnetic field on the structure of a white dwarf has been studied by \citet{Das2013, Das2014}. They showed that the inclusion of an interior huge magnetic field (of up to $\sim 10^{18}$~G), for the case of static white dwarfs, would increase the Chandrasekhar mass limit up to $\approx 2.58 M_{\odot}$. However, the detailed analyses performed by \citet{Coelho2014,2018mgm..conf.4363M} on the microscopic and macroscopic stability of such objects demonstrated that those masses are not attainable. \citet{Paret2015, Terrero2015} also calculated the effect of the magnetic field in white dwarfs. They showed that it was not possible to obtain stable magnetized white dwarfs with super-Chandrasekhar masses because the values of the surface magnetic field needed for them were higher than $10^{13}$~G. Thus \citet{Paret2015, Terrero2015} concluded that highly magnetized super-Chandrasekhar white dwarfs should not exist. Similar stability analyses were carried out by \citet{Chatterjee2017}. They concluded that strongly magnetized super-Chandrasekhar white dwarfs could not be  totally excluded from current theoretical considerations, though there are not any observational evidences.

The study of the structure of white dwarfs in the presence of spatially varying internal magnetic field that yields surface magnetic field $\sim 10^{13}$~G need further investigations \citep{Chamel2013}. This is, however, out of the scope of the current work. We constrain ourselves here to a surface dipole magnetic field up to the maximum observed values $\sim 10^{9}$~G \citep{2015MNRAS.446.4078K}. This field is involved only as the mechanism to slow down uniformly rotating white dwarfs. Therefore, we do not consider the influence of the magnetic field to the structure of white dwarfs.

Our paper is organized as follows: in section \ref{sec:2}, we discuss about WD equation of state, stability conditions and spin-up and spin-down effects; in section \ref{sec:4} we estimate the lifetime of SCWDs  at zero temperature; in section \ref{sec:5} we show the compression of the WD during its evolution; and in section \ref{sec:6} we consider the central temperature evolution for isothermal WD cores. Finally, we summarize our results in section \ref{sec:7}, discuss their significance, and draw our conclusions.

\section{WD equation of state and stability}\label{sec:2}

In this work, we will consider WDs surface magnetic field from $10^6$ to $10^9$~G, that are the highest magnetic field strengths  observationally  inferred with  polarimetry and Zeeman spectroscopy from WDs  data  \cite[see e.g.][]{2000PASP..112..873W}. If one assumes that the magnetic field inside the star is uniform for a WD with mass $M_{\rm wd}=1.44 M_\odot$ and radius $R_{\rm wd}=3000$~km \citep{2013ApJ...762..117B},  its  magnetic energy is about $E_m \approx (4\pi R_{\rm wd}^3/3)(B^2/8\pi)\sim 10^{39}-10^{45}$~erg, much smaller than its  gravitational energy $E_g\approx GM_{\rm wd}^2/R_{\rm wd}\sim 10^{50}$~erg. Thus, for the magnetic field strengths we are considering, the contribution of the magnetic energy to the structure equations of the star can be neglected.

It is also well known that a strong magnetic field will modify the equation of state of the matter due to the Landau quantization. However, even for magnetic field closed to the  critical value, which is defined by  $B_c=m_e^2c^3/\hbar e\approx 4.41 \times 10^{13}$~G, the Landau quantization of the electron gas has an negligible effect on the global properties of WDs as on its mass-radius relation \citep{2014MNRAS.445.3951B,2015PhRvD..92b3008C,2016MNRAS.456.3375B,Chatterjee2017}.

Although we  model  magnetized WDs, the unmagnetized description of the WD EoS and the WD structure will be a correct approximation, for the purposes of this article. Then, we follow in this work the treatment of \citet{2013ApJ...762..117B}. It is worth mentioning that for high magnetic fields a self-consistent study must be conducted, taking into account the effects of the magnetic field in the WD equation of state and structure equations \cite[see e.g.][]{Chatterjee2017}. For example, the  spherical symmetry of the star will be broken and the stability of a star will be limited by microphysical processes affected also by the magnetic field \citep{Coelho2014}. However, this issue is not considered here.

In this work we compute general relativistic configurations of uniformly rotating WDs within Hartle's formalism \citep{1967ApJ...150.1005H,2011IJMPE..20..136B,Terrero2017,2017IJMPS..4560025T,2018IJMPD..2750016T} and use the relativistic Feynman-Metropolis-Teller equation of state \citep{2011PhRvD..84h4007R,2011PhRvC..83d5805R} for WD matter. This equation of state generalizes the traditionally-used ones by \citet{chandrasekhar31} and  \citet{1961ApJ...134..669S}. A detailed description of this equation of state and the differences with respect to other approaches are given by \citet{2011PhRvD..84h4007R}.

The stability of uniformly rotating WDs has been analyzed taking into account
the Keplerian sequence (mass-shedding limit), inverse $\beta$-decay
instability, and secular axisymmetric instability by
\citet{2013ApJ...762..117B}. The mass-central density diagram of rotating WDs
composed of pure $^{12}$C for various angular momentum $J$=constant and angular
velocity $\Omega$=constant sequences has been illustrated in Fig~6 (Left panel)
by \citep{2013ApJ...762..117B}. In addition, there we have included the
critical density for pycnonuclear instability (in this case C+C reaction at
zero temperature) with reaction time scales $10^5$ and $10^{10}$~years. For
$^{12}$C WD, the maximun mass of the static configurations is $M_{\rm
max}^{J=0}=1.386~ M_\odot$ while for the uniformly rotating stars is $M_{\rm
max}^{J\neq 0}=1.474~M_\odot$.

\section{Super-Chandrasekhar WD lifetime}\label{sec:4}

We are interested in estimating the evolution of magnetized WDs when they are losing angular momentum via the magnetic dipole braking. It is well known that this braking mechanism needs, for a rotating magnetic dipole in vacuum, that the magnetic field is misaligned with the rotation axis. It has been shown that such WDs with misaligned fields could be formed in the merger of double degenerate binaries, and the degree of misalignment depends on the difference between the masses of the WD components of the binary \citep{2012ApJ...749...25G}.

Particularly interesting are the SCWDs which are stable by virtue of their angular momentum. Thus, by loosing angular momentum, they will evolve toward one of the aforementioned instability borders. This is the reason why, for example, magnetic braking of SCWDs has been invoked as a possible mechanism to explain the delayed time distribution of type Ia supernovae \citep{ilkov2012}: a type Ia supernova can be delayed for a time typical of the spin-down time scale $\tau$ due to magnetic braking, providing the result of the merging process of a WD binary system is a magnetic SCWD rather than a sub-Chandrasekhar one. It is important to recall that in such a model it is implicitly assumed that the fate of the WD is a supernova explosion instead of gravitational collapse. The competition between these two possibilities is an important issue which deserves to be further analyzed but it is out of the scope of this work.

Since SCWDs spin-up by angular momentum loss the reference to a ``spin-down'' time scale for them can be misleading. Thus, we prefer hereafter to refer the time the WD spend in reaching an instability border (either mass-shedding, secular or inverse $\beta$ instability, see \citep{2013ApJ...762..117B}).

\begin{figure}
\centering
\includegraphics[width=9.3 cm,clip]{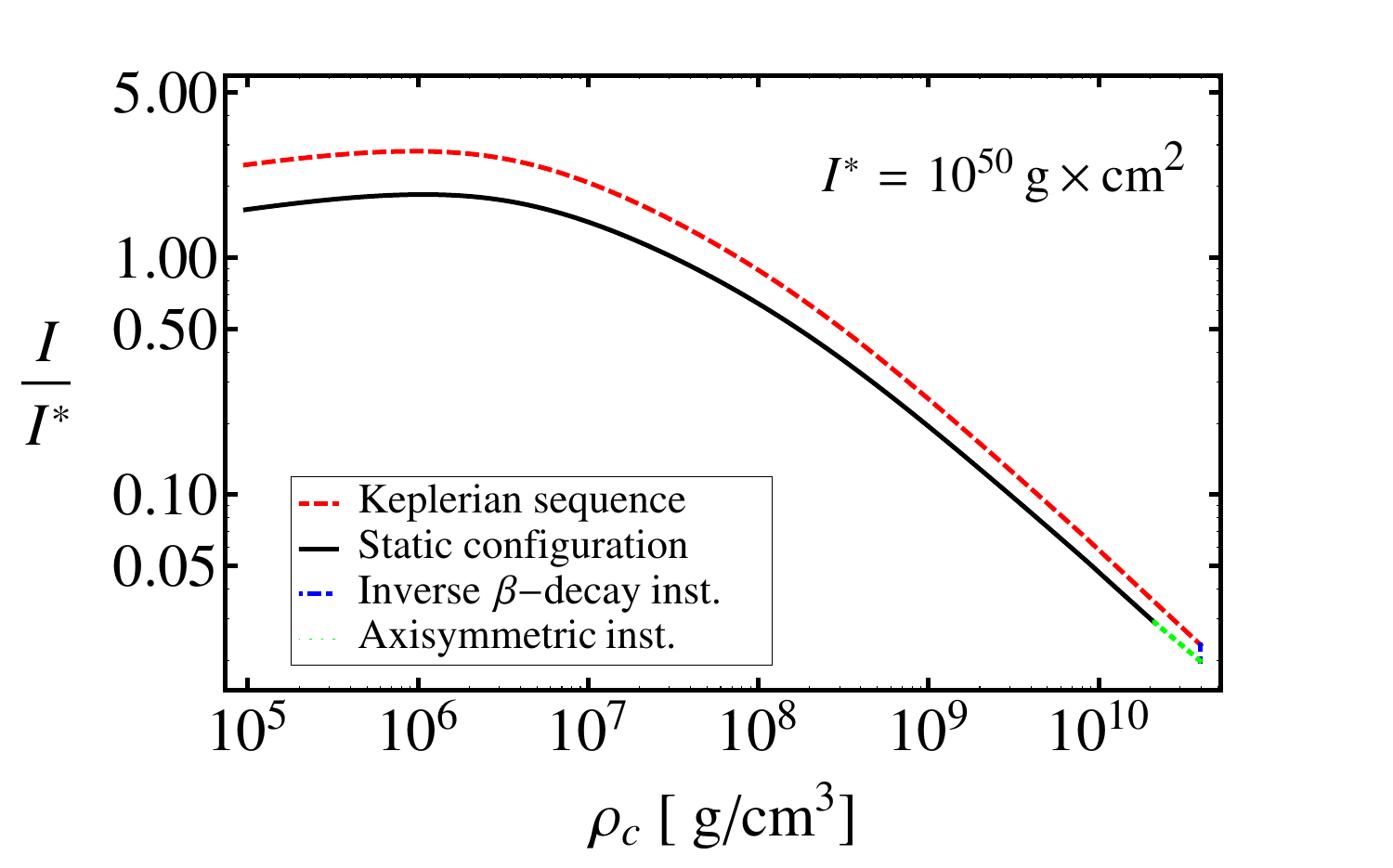}
\caption{(Color online) Moment of inertia versus central density.}\label{fig:irho}
\end{figure}

As we have shown, the evolution track followed by a SCWD depends strongly on the initial conditions of mass and angular momentum, as well as on the nuclear (chemical) composition and the evolution of the moment of inertia. Clearly, the assumption of fixed moment of inertia leads to a lifetime scale that depends only on the magnetic field's strength. A detailed computation will lead to a strong dependence on the mass of the SCWD, resulting in a two-parameter family of lifetime $\tau=\tau(M,B)=t$. Indeed, we see from Fig.~\ref{fig:irho} that the moment of inertia even for a static case is not constant since it is a function of the central density. In the rotating case it is a function of both central density and angular velocity.

Thus, we have estimated the characteristic lifetime relaxing the constancy of the moment of inertia, radius and other parameters of rotating WDs. In fact we show here that all WD parameters are functions of the central density and the angular velocity. We used the  original form of the lifetime formula
\begin{equation}\label{eq:lifetime}
dt=-\frac{3}{2}\frac{c^3}{ B^2}\frac{1}{R^6 \Omega^3} dJ,
\end{equation}
where $c$ is the speed of light, $B$ is the magnetic field intensity in Gauss, $J$ is the angular momentum, $\Omega$ is the angular velocity and $ R$ is the mean radius calculated along the specific constant-rest-mass sequences.

Hence, we performed more refined analyses with respect to \citet{ilkov2012} by taking into consideration all the stability criteria, with the only exception of the instabilities related to nuclear reactions (either pynonuclear or enhanced by finite temperature effects) which we will consider in Sec.~\ref{sec:6}.

\begin{figure}
\centering
\includegraphics[width=\hsize,clip]{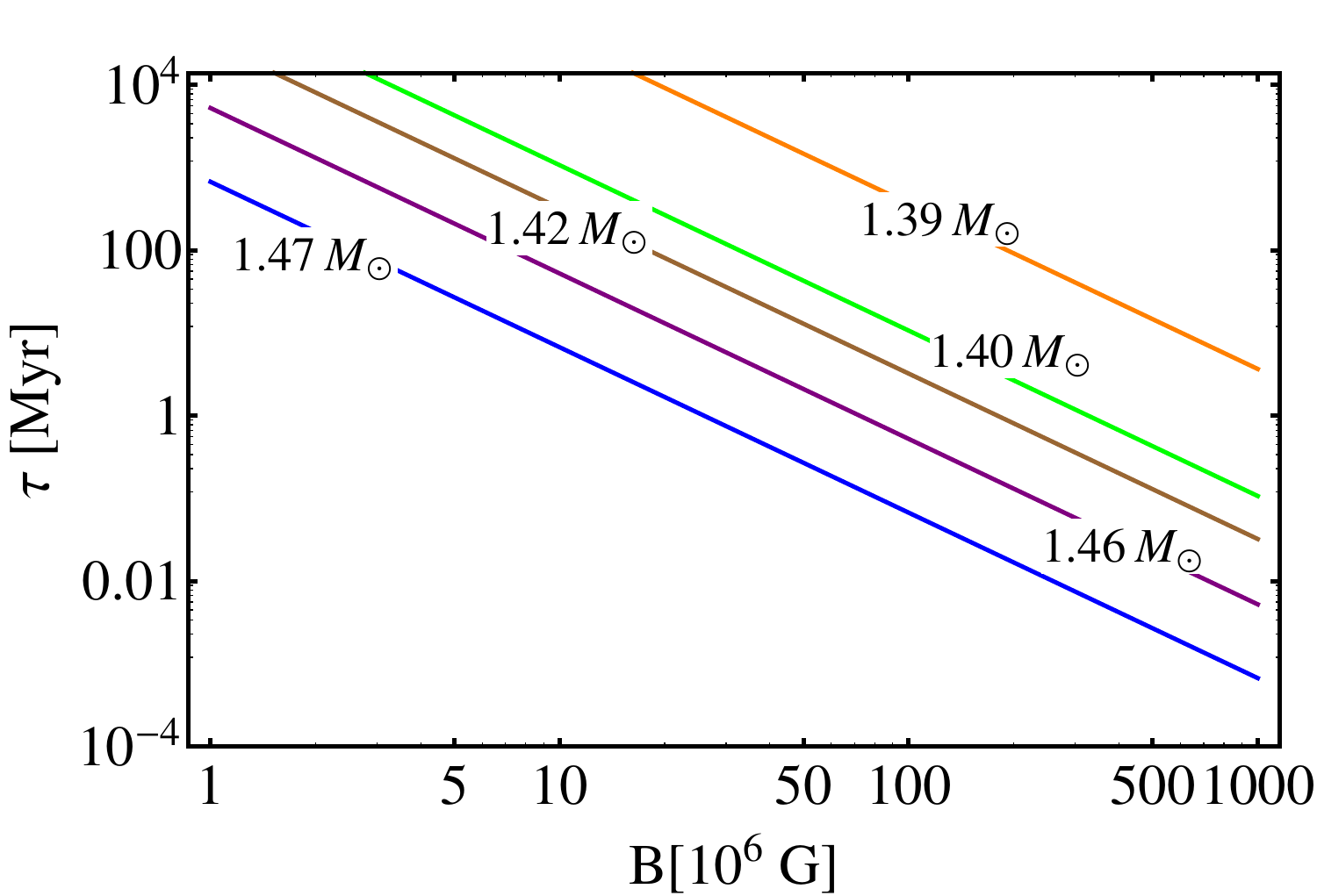}
\caption{(Color online) Characteristic life time $\tau$ in Myr versus WD surface magnetic field $B$ in Gauss for rotating $^{12}$C  WDs.}\label{fig:lifetime}
\end{figure}

The characteristic lifetime $\tau$ as a function of the surface magnetic field for given sets of constant mass sequences is demonstrated in Fig.~\ref{fig:lifetime}. Here each straight line corresponds to a certain fixed constant mass sequence. By choosing one value or the whole range of the magnetic field intensity one can estimate the lifespan of a SCWD for a given mass. One can see that the higher the magnetic field, the shorter the lifetime of the rotating SCWD. Correspondingly, a more massive WD will have a shorter lifespan and vice versa.

Another interesting representation of  the characteristic lifetime $\tau$ as a function of WD mass in units of $M^{J=0}_{max}$ for zero temperature uniformly rotating $^{12}$C WDs has been considered in Fig.~3 by \citet{2014JKPS...65..855B}. There, one can see how for a fixed magnetic field value, the lifetime has a wide range of values as a function of the mass, being inversely proportional to the latter. The time scales of \citet{ilkov2012,2013MNRAS.431.2778K} appear to be consistent with the one in Fig.~\ref{fig:lifetime} only for the maximum mass value $M/M^{J=0}_{max}\approx 1.06$.

\section{Induced compression}
\label{sec:5}

To investigate the evolution of isolated white dwarfs with time we made use of Eq.~\eqref{eq:lifetime} and modified it depending on what parameter we are interested in (see \citep{2014IJMPKazNU, 2017mgm} for details). Consequently, we adopt two cases: 1) when the magnetic field is constant and 2) when magnetic flux is conserved
\begin{eqnarray}\label{eq:conts_B}
B&=&B_0 ,\\
B&=&B_0\frac{ R_0^2}{{ R}^2},\label{eq:flux}
\end{eqnarray}
where $B_0$ is the surface dipole magnetic field corresponding to the initial value of $B$ at $t=0$, ${ R_0}$ is the mean radius corresponding to the initial values of $ R$ at $t=0$. Plugging $B$ in Eq.~(\ref{eq:lifetime}) we obtain two separate equations that describe the evolution of a considered parameter with time taking into account two cases when magnetic field is constant and when magnetic flux is conserved. The behavior of the central density, mean radius and the magnetic field intensity as a function of the characteristic lifetime for both cases have been investigated by \citet{2014IJMPKazNU}. Similar analyses have been performed for the angular velocity as a function of the characteristic lifetime by \citet{2017mgm}.

\begin{figure}
\centering
\includegraphics[width=\hsize,clip]{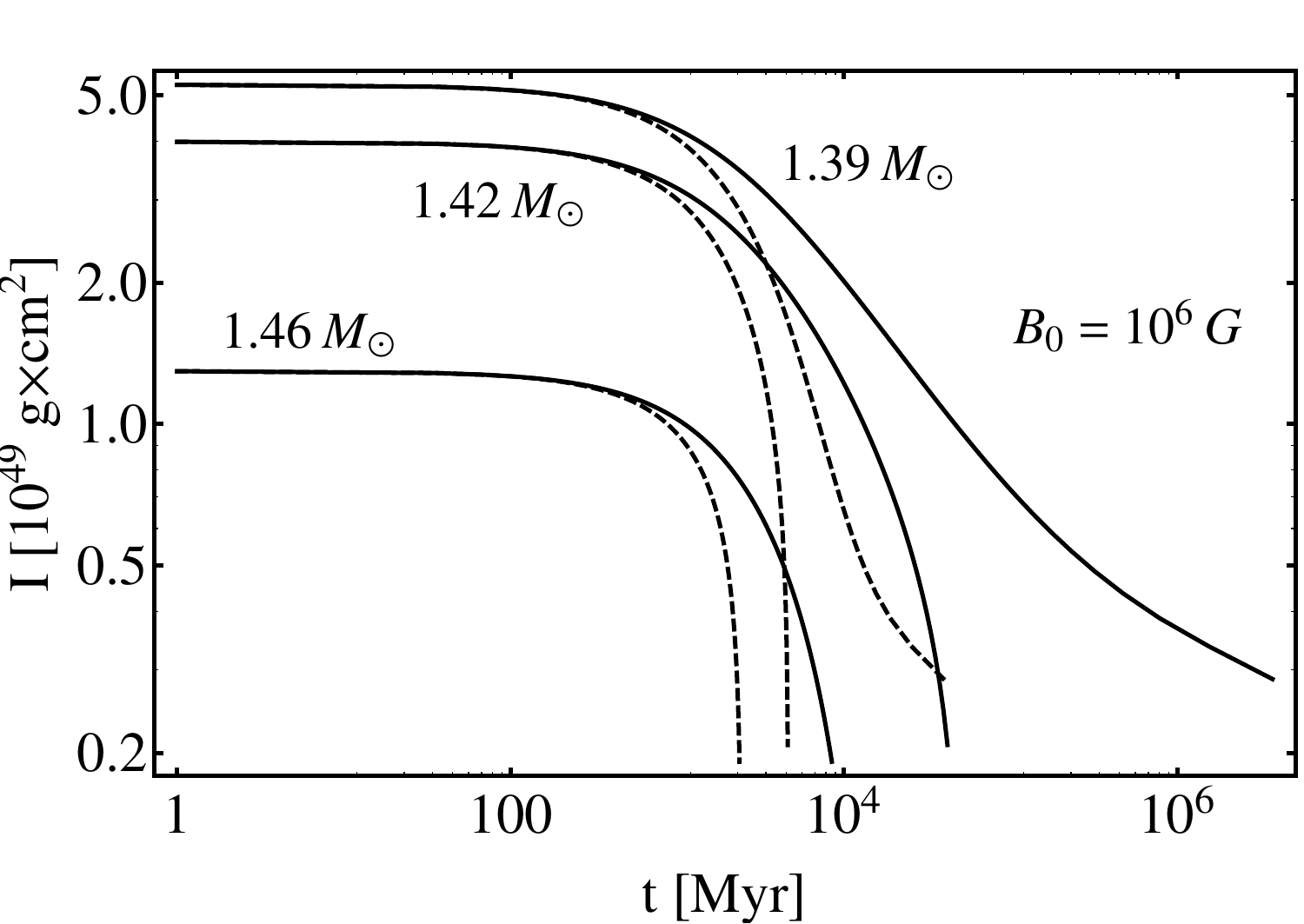}
\caption{Moment of inertia versus time.  Solid curves indicate the evolution path when the magnetic field is constant and dashed curves indicate the evolution path when the magnetic flux is conserved with $B_0=10^6$ G for selected constant mass sequences.}\label{fig:iti}
\end{figure}

\begin{figure}
\centering
\includegraphics[width=\hsize,clip]{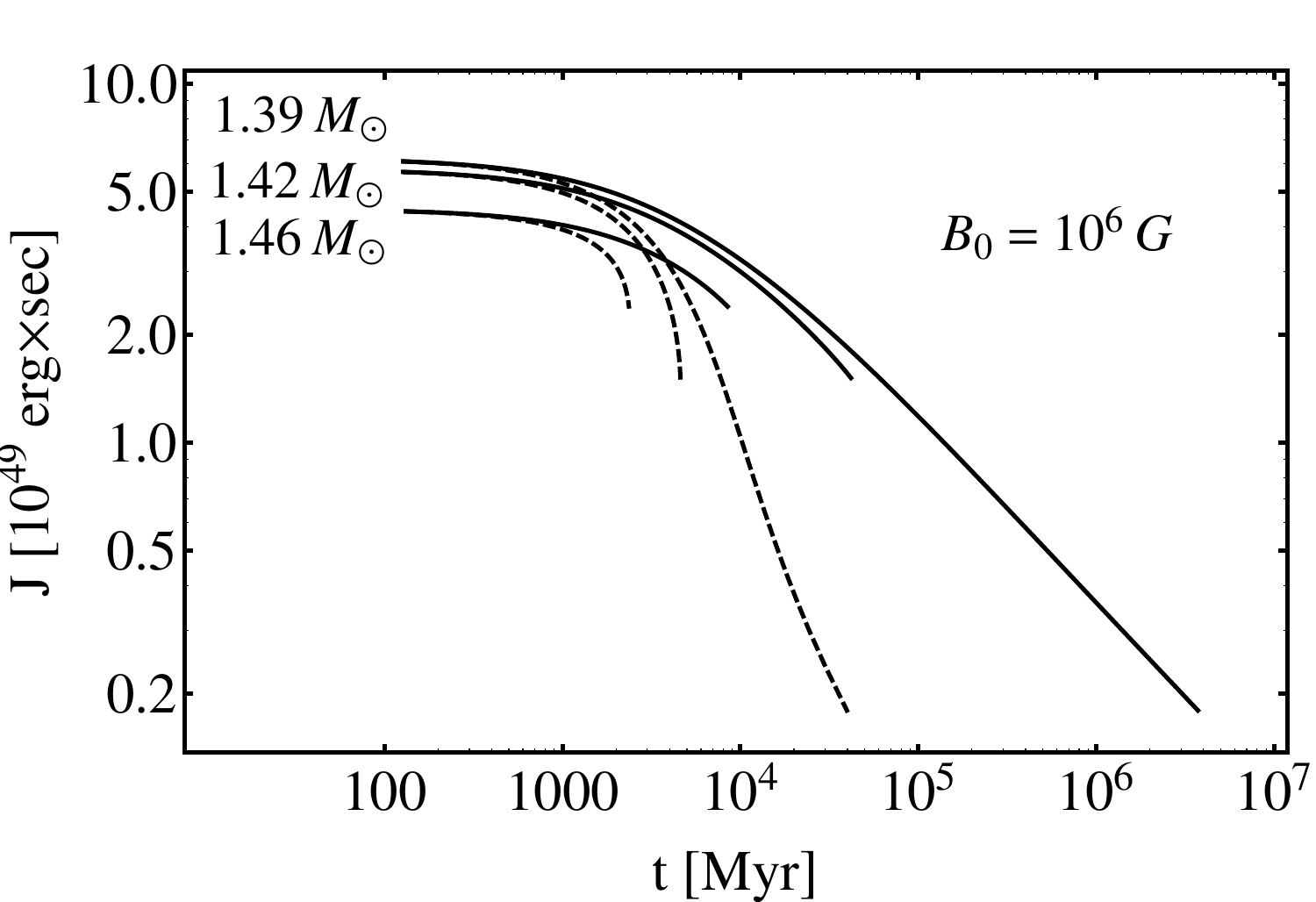}
\caption{Angular momentum versus time. Solid curves indicate the evolution path when the magnetic field is constant and dashed curves indicate the evolution path when the magnetic flux is conserved with $B_0=10^6$ G for selected constant mass sequences.}\label{fig:jti}
\end{figure}

For isolated rotating white dwarfs the rest mass remains unchanged in their entire evolution time and hence the moment of inertia of WDs in Fig.~\ref{fig:iti} has a direct correlation with the mean radius (see \cite{2017mgm}). Since the angular momentum is lost by magnetic dipole braking in Fig.~\ref{fig:jti}  we see for larger masses it decreases even faster than for smaller masses.

Overall, as one can see from the plots above the isolated WDs regardless of their masses due to magnetic dipole braking will always lose angular momentum (see e.g. Fig.~\ref{fig:jti}). Therefore WDs will tend to reach a more stable configuration by increasing their central density and by decreasing their mean radius (see e.g. \cite{2017mgm}). Ultimately, super-Chandrasekhar WDs will spin-up and sub-Chandrasekhar WDs spin-down. As for the WDs having masses near the Chandrasekhar mass limit, they  will experience both spin up and spin down epochs at certain time of their evolution \citep{2013ApJ...762..117B}.

\begin{figure*}
\centering
\includegraphics[width=0.49\hsize,clip]{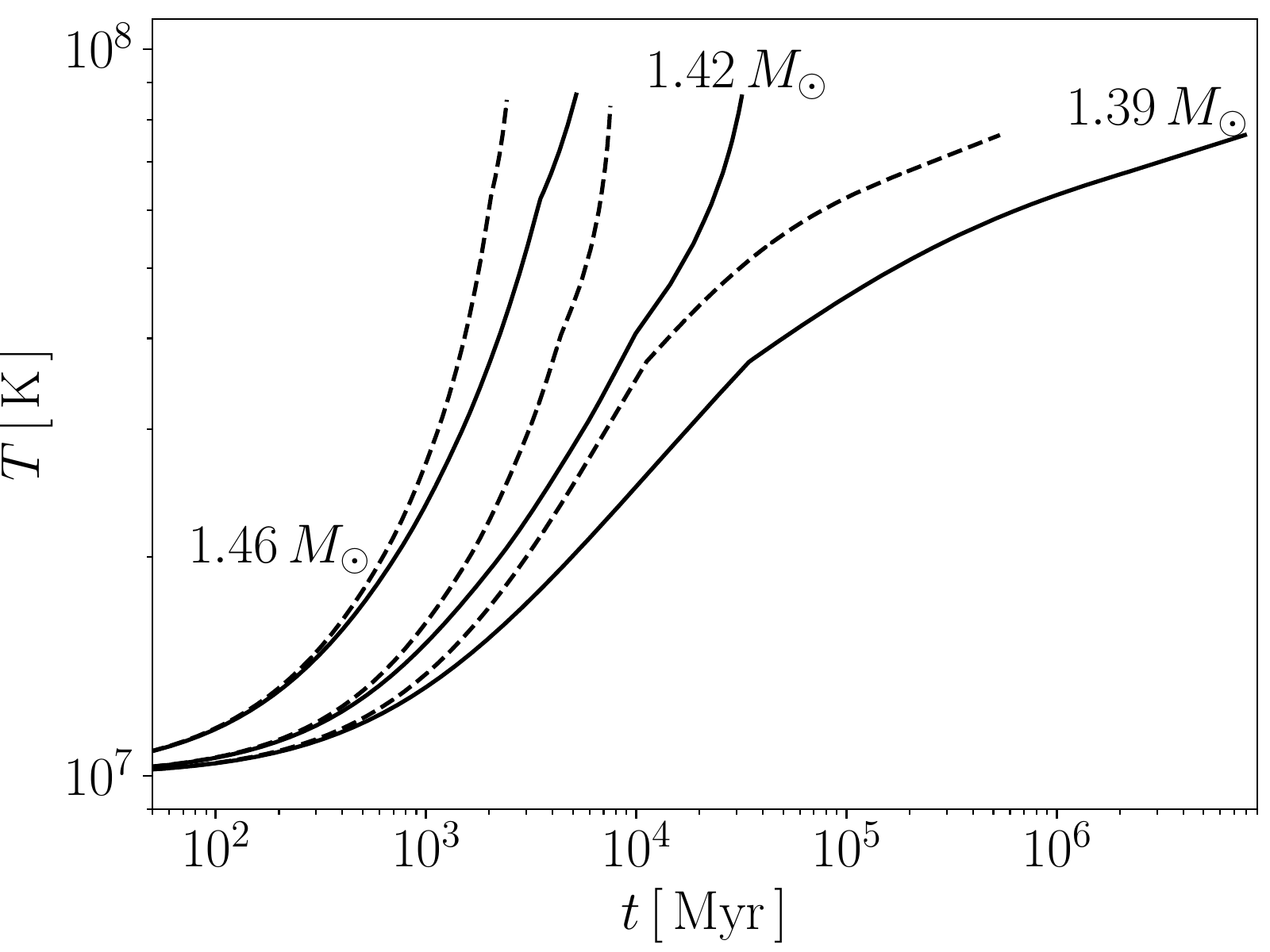} \includegraphics[width=0.49\hsize,clip]{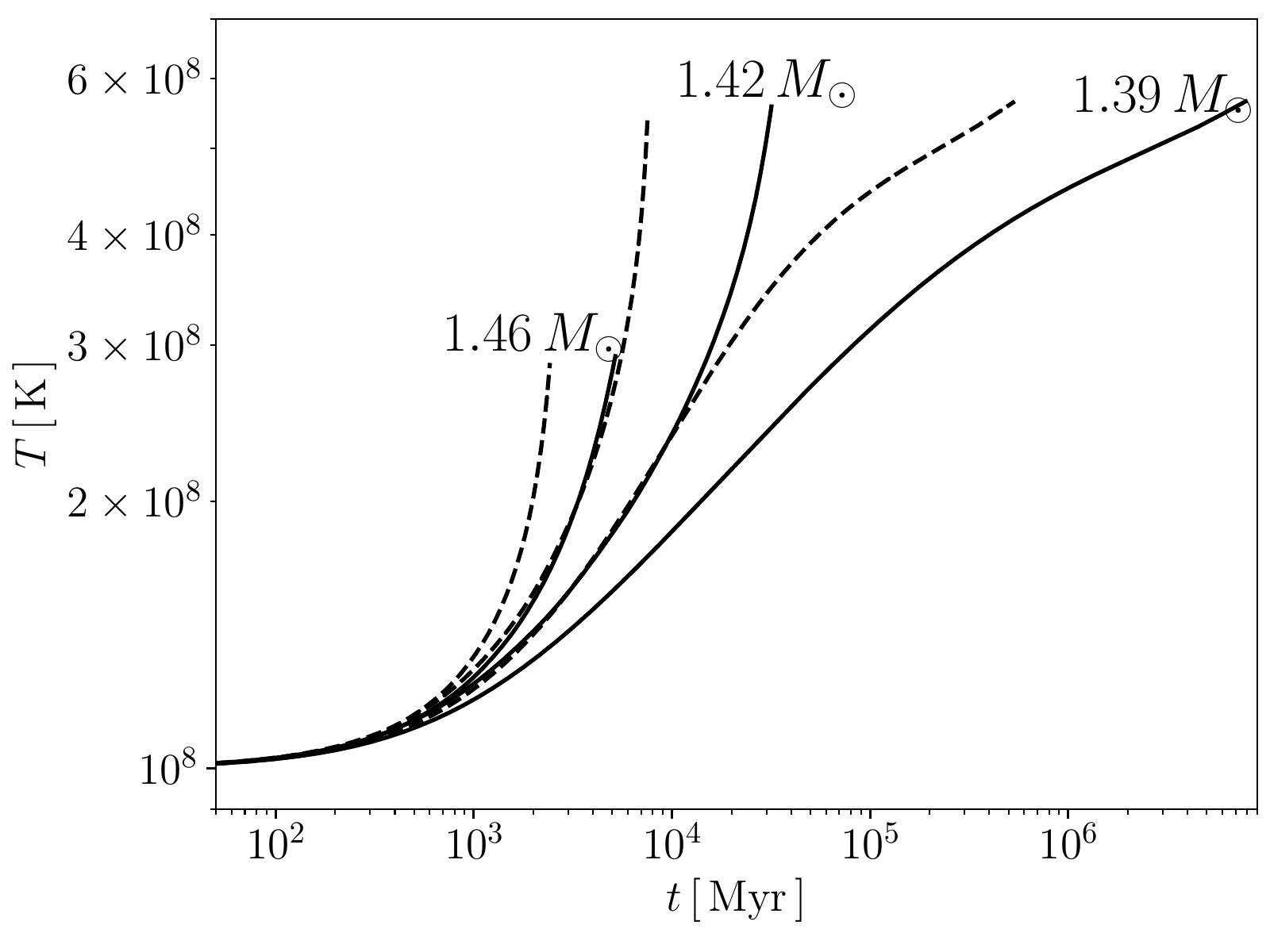}
\caption{Central temperature versus time for constant mass sequences. Solid curves indicate the evolution path when the magnetic field is constant and dashed curves indicate the evolution path when the magnetic flux is conserved with $B_0=10^6$ G. Left panel:  $T(0)=10^{7}$~K. Right panel: $T(0)=10^{8}$~K.}\label{fig:Temi_7}
\end{figure*}

%
\section{Central Temperature evolution}\label{sec:6}

%
In this section, we model a SCWD as an isothermal core and follow its thermal evolution by solving the equation of energy conservation:
\begin{equation}\label{eq:cons_energy}
\frac{dL}{dm}=\epsilon_{\rm nuc}-\epsilon_{\nu}+T\dot{s}
\end{equation}
where $L$ is the luminosity, $\epsilon_{\rm nuc}$ is the nuclear reactions energy release per unit mass, $\epsilon_{\nu}$ is the energy loss per unit mass by the emission of neutrinos, $T$ is the temperature and $s$ is the specific entropy. The last term of Eq.~(\ref{eq:cons_energy}) can be written as:
\begin{equation}\label{eq:dotS}
T\dot{s}=c_{\rm v}\dot{T}-\left[\frac{P}{\rho^2}-\left(\frac{\partial u}{\partial\rho}\right)_T\right]\dot{\rho}
\end{equation}
where $u$ is the internal energy and $c_{\rm v}$ is the heat capacity at constant volume. For the latter, we used the analytic fits of \citet{1998PhRvE..58.4941C,2000PhRvE..62.8554P}. The heat capacity is calculated from the Helmholtz free energy assuming a fully ionized plasma consisting of point-like ions immersed in an electron background. The plasma is characterized by the Coulomb coupling parameter, defined as the ratio between the potential energy and the thermal energy: $\Gamma=(Ze)^2/(\kappa_BTa)$, where $a=(4\pi/3\, n_i)^{-1/3}$ is the mean inter-ion distance and $n_i$ is the ion number density. At $\Gamma\lesssim 1$ the ions behave as a gas, at $\Gamma>1$ as a strongly coupled Coulomb liquid, while crystallization occurs at $\Gamma=\Gamma_m\approx 175$. The quantum effects are taken into account when $T_p\ll\hbar \omega_p/\kappa_B$, where $\omega_p =(4\pi Z^2e^2 n_i/m_i)^{1/2}$ is the ion plasma frequency.

First, we integrate Eq.~\eqref{eq:cons_energy} with Eq.~\eqref{eq:dotS} neglecting the nuclear burning and neutrino emission processes. Fig.~\ref{fig:Temi_7} shows the evolution of the WD central temperature for constant mass sequences ($M_{\rm WD}=1.39, 1.42$ and $1.46\, M_\odot$) with two different initial temperature: $T(0)=10^7$~K and $T(0)=10^8$~K, respectively. They were obtained by solving simultaneously Eq.~(\ref{eq:lifetime}) and:
\begin{equation}\label{eq:eq1}
dT = \frac{P}{c_{\rm v}(\rho,T)\rho}\frac{\partial\, {\rm log}\, \rho}{\partial J}\,dJ\,.
\end{equation}
For the magnetic field evolution, we have adopted the two cases of the previous section: constant surface magnetic field (see Eq.~\eqref{eq:conts_B}) and constant magnetic flux (see Eq.~\eqref{eq:flux}) with $B_0=10^6$~G. It is obvious that regardless of the initial temperature, the WD configuration heats up while it is compressing, as it is seen in Fig.~\ref{fig:Temi_7}.

In particular case, when $T(0)=10^7$~K (left panel of Fig.~\ref{fig:Temi_7}), there is a phase transition from solid to liquid state while the system is heating. This could change the thermal evolution due to the latent heat and will depend on the abundance of chemical species (see e.g. \citep{1997ApJ...485..308I,1997IAUS..189..381C}). However, when $T(0)=10^8$~K (right panel of Fig.~\ref{fig:Temi_7}), the configuration of the WD has a Coulomb parameter $\Gamma<175$, i.e. the matter is in liquid state along its entire evolution.

In order to introduce the nuclear burning and the neutrinos emission process we have considered that for the rotating $^{12}$C WDs the thermonuclear energy is essentially released by two nuclear reactions:
\begin{eqnarray}
^{12}{\rm C}+^{12}{\rm C}\,&\rightarrow&\,^{20}{\rm Ne}+\alpha+4.62\,{\rm MeV},\\
^{12}{\rm C}+^{12}{\rm C}\,&\rightarrow&\,^{23}{\rm Na}+{\rm p}+2.24\,{\rm MeV},
\end{eqnarray}
with  nearly  the  same  probability. For the  carbon  fusion reaction rate, we used the fits obtained by \citet{2005PhRvC..72b5806G} and for the neutrino energy losses we used the analytical fits calculated by \citet{1996ApJS..102..411I}, which account for the electron-positron pair annihilation ($e^{-}e^{+}\rightarrow\nu\bar{\nu}$), photo-neutrino emission ($e+\gamma\rightarrow e\nu\bar{\nu}$), plasmon decay ($\gamma\rightarrow\nu\bar{\nu}$), and electron-nucleus bremsstrahlung [$e(Z,A)\rightarrow e(Z,A)\nu\bar{\nu}$].

\begin{figure*}
\centering
\includegraphics[width=0.49\hsize,clip]{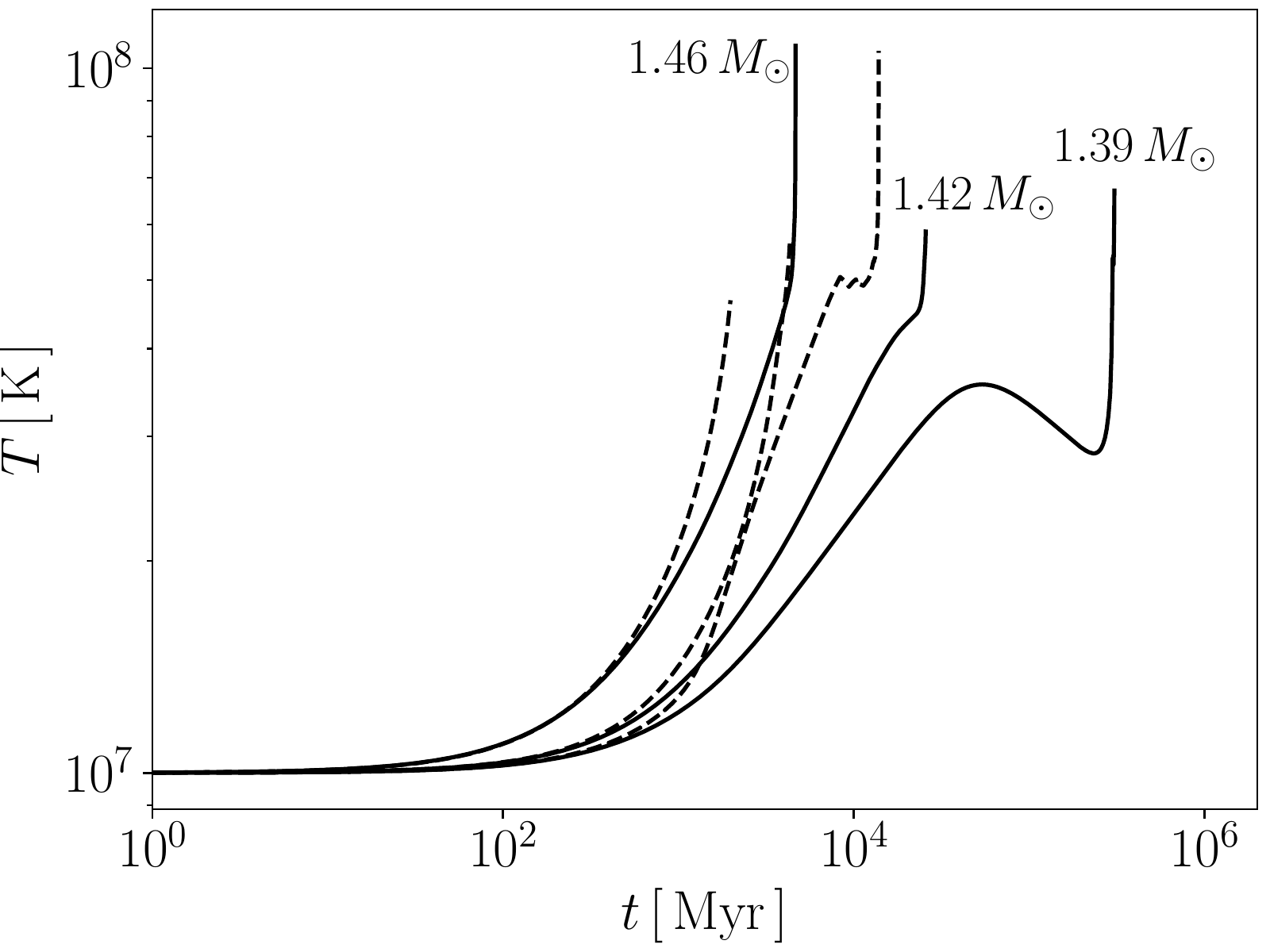} \includegraphics[width=0.49\hsize,clip]{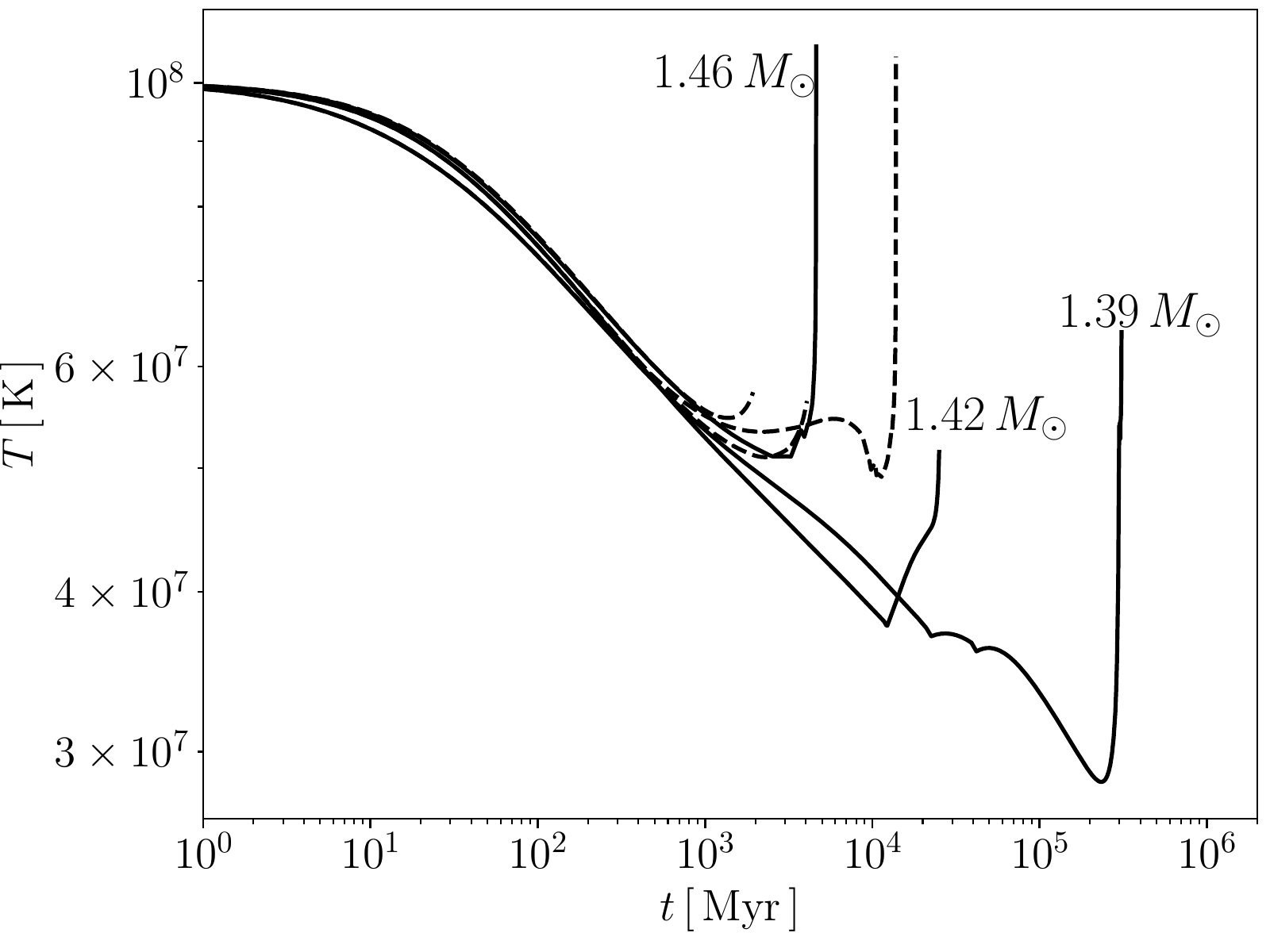}
\caption{Central temperature versus time introducing carbon fusion energy release and neutrino energy losses. Solid curves indicate the evolution path when the magnetic field is constant and dashed curves indicate the evolution path when the magnetic flux is conserved with $B_0=10^6$ G for fixed constant mass sequences. Left panel:  $T(0)=10^{7}$~K. Right panel: $T(0)=10^{8}$~K.}\label{fig:Teme_7}
\end{figure*}

The central temperature evolution is shown in Fig.~\ref{fig:Teme_7} when the carbon fusion energy release and the neutrinos energy loses are taken into account. This was done by solving:
\begin{equation}\label{eq:eq2}
dT = \frac{P}{c_{\rm v}(\rho,T)\rho}\frac{\partial \, {\rm log}\,\rho}{\partial J}\,dJ + \frac{\epsilon_{\rm nuc}-\epsilon_\nu}{ c_{\rm v}(\rho,T)}\, dt
\end{equation}
At the beginning, when $T(0)=10^7$~K (left panel of Fig.~\ref{fig:Teme_7}), the configuration heats via the energy release from the carbon fusion and the compression of the configuration. However, when $T(0)=10^8$~K (right panel of Fig.~\ref{fig:Teme_7}), the energy losses from the neutrino emission process is the dominant process and it cools the configurations.

\begin{figure}
\centering
\includegraphics[width=\hsize,clip]{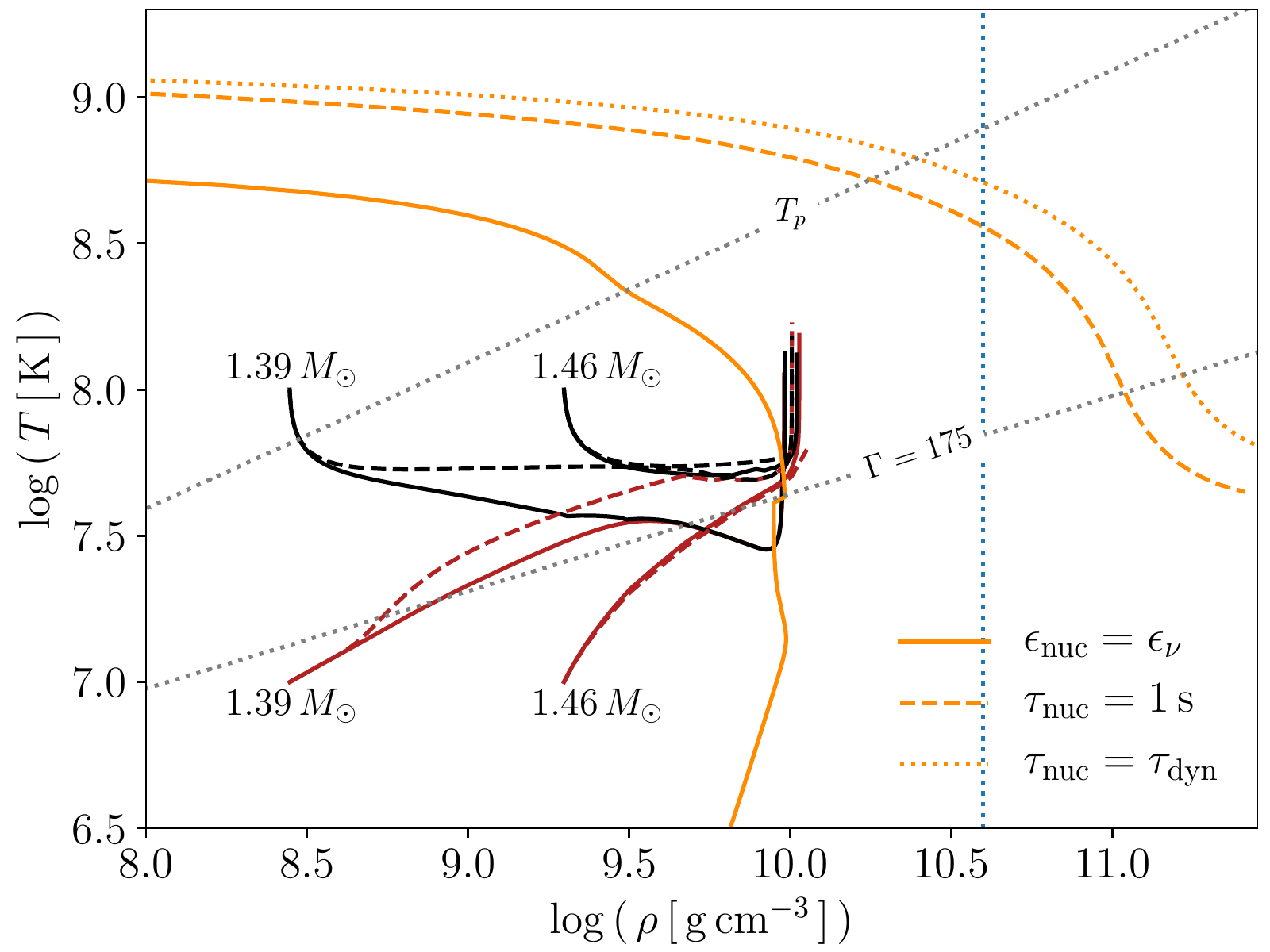}
\caption{(Color online) The evolution track in the central density and central temperature plane  with carbon fusion energy release and neutrino energy losses. We also show the carbon-ignition line (solid orange line), the line along which $\tau_{\rm nuc}=1$~s and $\tau_{\rm nuc}=\tau_{\rm dyn}$.   the crystallization line labeled as $\Gamma=175$, and the plasma temperature line defined as $\kappa_BT_p=\hbar\omega_p$.}\label{fig:Tem_ign}
\end{figure}

Fig.~\ref{fig:Tem_ign} shows the evolutionary path of some constant mass sequences in the central temperature and central density plane. It also exhibits the lines where the neutrino emissivity equals the nuclear energy release (i.e. carbon-ignition line: $\epsilon_{\rm nuc}=\epsilon_\nu$) and  the lines along which the characteristic time of nuclear reactions, $\tau_{\rm nuc}$, equals one second (1 s) and equals the dynamical timescale, $\tau_{\rm dyn}$.  The dynamical timescale $\tau_{\rm dyn}$ is defined as:
\begin{equation}\label{eq:dynamical_time}
\tau_{\rm dyn} = \frac{1}{\sqrt{24\pi G \rho}}\, ,
\end{equation}
and the characteristic time of nuclear reactions is:
\begin{equation}
\tau_{\rm nuc}=\frac{\epsilon_{\rm  nuc}}{\dot{\epsilon}_{\rm nuc}}=c_p\left(\frac{\partial\epsilon_{\rm nuc}}{\partial T}\right)^{-1}
\end{equation}
As seen in Fig.~\ref{fig:Tem_ign}, all the configurations cross  the carbon-ignition line. When this happens, the star will be heated evolving to a state at which the fusion reactions becomes instantaneous ($\tau_{\rm nuc}<\tau_{\rm dyn}$), possibly leading to a supernova.
For some constant mass  sequence, the configuration crosses the crystallization line, labelled as $\Gamma=175$ in Fig.~\ref{fig:Tem_ign}. When $T(0)=10^8$~K, it happens just for $M=1.39\,M_\odot$, while for lower initial temperature it  happens for all the constant mass sequences considered here.

\begin{table*}
\centering
\caption{Total WD Time Evolution ($B_0=10^6$~G): the first four columns correspond to the WD total mass, $M$, WD initial angular velocity, $\Omega_0$,  initial dynamical timescale, $\tau_{\rm dyn,0}$ and initial central temperature, $T(0)$, respectively. The following three columns correspond to the time needed by the WD to reach  an instability boundary, $\tau$, and to reach the carbon-ignition line when only the compression of the star is considered, $\Delta\tau_\rho$, and when the energy release from the carbon fusion and neutrino emission is introduced, $\Delta\tau_{\rm CC}$. First when the surface magnetic field is constant and then when the magnetic flux is conserved. }\label{tab:times}
\begin{tabular}{cccc|ccc|ccc}

  \multicolumn{4}{c|}{WD Parameters }  & \multicolumn{3}{c|}{Constant Magnetic Field}  & \multicolumn{3}{c}{Constant Magnetic Flux } \\ \hline
  \hline
  $M$ & $\Omega_0$ & $\tau_{\rm dyn,0}$ & $T(0)$ & $ \tau $ & $\Delta \tau_\rho$ & $\Delta \tau_{\rm CC}$ & $\tau$ & $\Delta \tau_\rho$  &$\Delta \tau_{\rm CC}$\\
 $M_{\odot}$ & s$^{-1}$ & $10^{-4}$~s & K & Myr &  Myr & Myr & Myr & Myr &  Myr  \\
 \hline \hline
 &  &  & $10^8$ &  & $3.68\times 10^4$ & $2.64\times 10^5$ &  &$6.54\times 10^3$   & $1.31\times 10^4$ \\
 $1.39$  & $1.59$ &$7.07$ & $10^7$ & $3.76 \times 10^6$  & $3.09\times10^5$  & $2.64\times 10^5$ & $3.99\times 10^{4}$ & $1.35\times 10^4$  &$1.31\times 10^4$  \\
  & & & $10^6$ & & $3.13\times10^5$ & $2.64\times 10^5$ &  & $3.54\times 10^4$  & $1.31\times 10^4$  \\
  \hline
  & &  & $10^8$ &  & $1.12\times 10^4$ & $2.53\times 10^4$&  &$3.46\times10^3$ & $4.28\times 10^3$ \\
1.42  & 1.94  & $42.4$& $10^7$ &  $4.28 \times 10^4$ & $2.54\times 10^4$  &$2.53\times 10^4$ & $4.59\times 10^{3}$ & $4.31\times10^3$ &  $4.28\times 10^3$\\
  & & & $10^6$ & & $2.57\times 10^4$  & $2.53\times 10^4$ & &$3.99\times10^3$  & $4.28\times 10^3$ \\
 \hline
  &  &  & $10^8$ &  &$2.83 \times 10^3$ & $4.39\times 10^3$&  & $1.91\times 10^3$ & $2.03\times 10^3$ \\
1.46  & 3.06 & $13.5$ & $10^7$ & $8.51 \times 10^3$ & $2.34 \times 10^3$ & $4.45\times 10^3$& $2.37\times 10^{3}$ &$2.24\times 10^3$  & $2.03\times 10^3$  \\
  & & & $10^6$ & & $4.53 \times 10^3$ & $4.45\times 10^3$&  & $2.03\times 10^3$ &  $2.03\times 10^3$ \\
\end{tabular}
\end{table*}

In Tab.~\ref{tab:times}, we compare the time needed by the configuration to reach an instability limit (mass-shedding, secular axisymmetric instability and/or inverse $\beta$ decay instability), $\tau$ (see Eq.~(\ref{eq:lifetime})), with the one needed to reach the carbon-ignition line when one just considers the compression of the star, $\Delta\tau_\rho$ (solving Eq.~(\ref{eq:eq1})), and when energy released from the carbon fusion and neutrino emission are introduced, $\Delta\tau_{\rm CC}$ (solving Eq.~(\ref{eq:eq2})), for the both magnetic field evolution: constant magnetic field and constant magnetic flux with $B_0=10^6$~G.

We have also specified the  mass of the configuration, $M$, the initial angular velocity, $\Omega_0$, the magnitude of the initial dynamical timescale, $\tau_{\rm dyn,0}$ as well as the initial central temperature, $T(0)$. In all cases the star arrives first to the ignition line then to the instability limit. This time difference is greater for less massive WDs than for more massive ones.

\begin{figure}
\centering
\includegraphics[width=\hsize,clip]{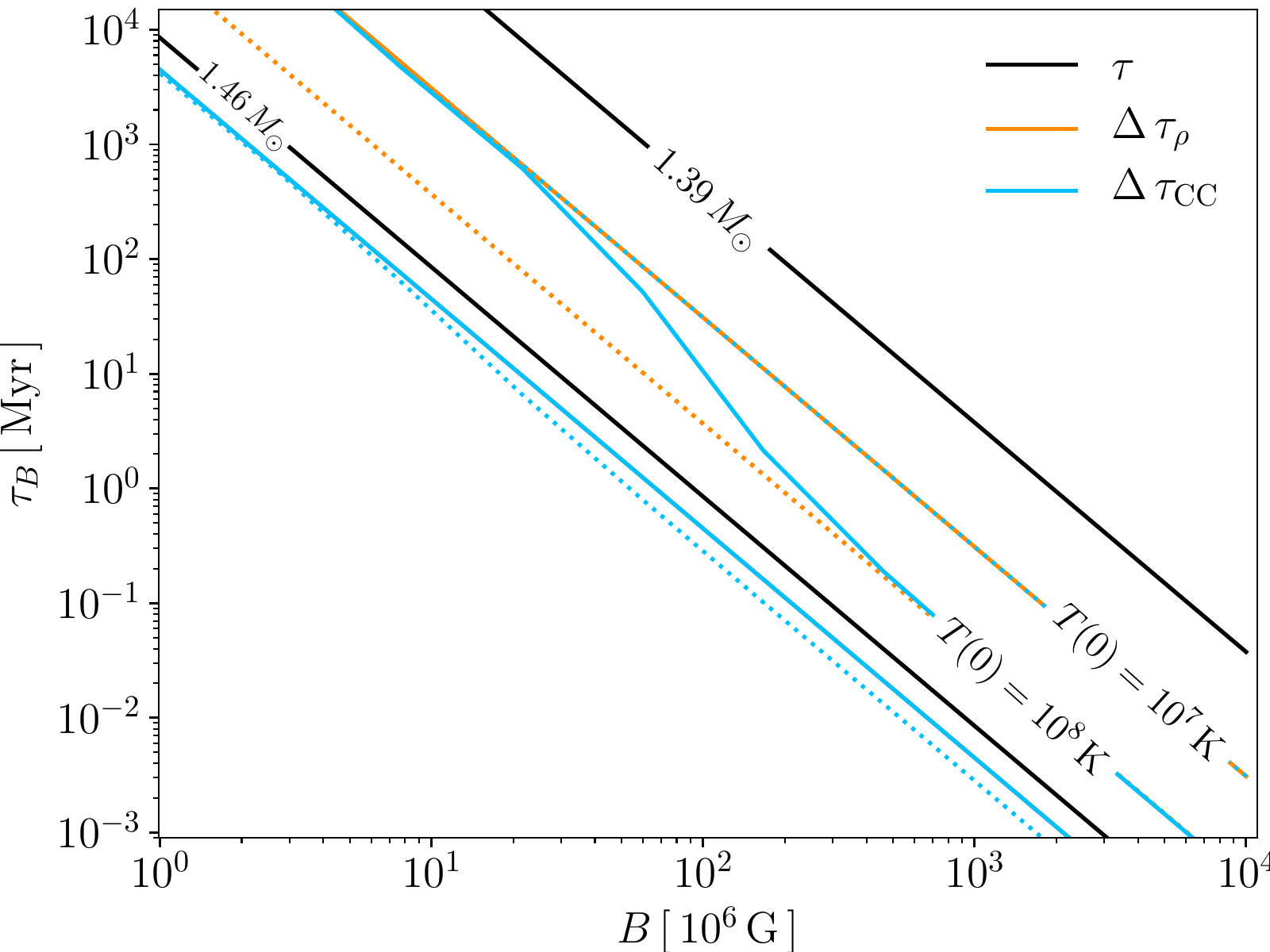}
\caption{(Color online) Characteristic times of the WD evolution as a function of the surface magnetic field. Here, it is compared with the total time needed by the WD configuration to reach an instability limit, $\tau$ (solid black lines),  the one needed to reach the carbon-ignition line when the compression of the star is considered, only, $\Delta\tau_\rho$ (orange lines), and when the energy release from the carbon fusion and neutrino emission is introduced, $\Delta\tau_{\rm CC}$ (blue lines).}\label{fig:timescale}
\end{figure}

Finally, in Fig.~\ref{fig:timescale}, we show the $\tau$, $\Delta\tau_\rho$ and $\Delta\tau_{\rm CC}$ as a function of the surface magnetic field strength when the field intensity is assumed to be constant along all the evolution, for different constant mass sequences and different initial temperature. The stronger surface  magnetic field the shorter the WD lifetime.

\section{Concluding Remarks}\label{sec:7}
%
We have investigated in this work the evolution of the WD structure while it loses angular momentum via magnetic dipole braking. We obtained the following conclusions:
\begin{enumerate}

\item
We have computed the lifetime of SCWDs as the total time it spends in reaching one of the following possible instabilities: mass-shedding, secular axisymmetric instability and inverse $\beta$ decay instability. The lifetime is inversely proportional both to the magnetic field and to the mass of the WD (see Fig.~\ref{fig:lifetime}).
\item
We showed how the parameters of rotating WDs evolve with time. For the sake the of comparison we considered two cases: a constant magnetic field and a varying magnetic field conserving magnetic flux. We showed that the WD is compressed with time. It turns out that, in the case of magnetic flux conservation, the evolution times are shorter than for the constant magnetic field, hence the WD lifetime.
\item
The time scales of \citet{ilkov2012,2013MNRAS.431.2778K} are consistent with the ones in Fig.~\ref{fig:lifetime} only for the maximum mass value $M/M^{J=0}_{max}\approx 1.06$.
\item
Whether or not the SCWD can end its evolution as a type Ia supernova as assumed e.g. by \citet{ilkov2012} is a question that deserves to be further explored. Here, we have computed the evolution of the WD central temperature while it losses angular momentum along constant mass sequence. In general,  the WD will increase its temperature while it is compressed. When carbon fusion reaction and neutrino emission cooling are considered, the evolution depends on the initial temperature. However, in all the cases studied, the configurations reach the carbon-ignition line. In this point the speed of the carbon fusion reactions  increases until reaching conditions for a thermonuclear explosion.

\item
We have also computed the time that the SCWDs need in reaching the carbon-ignition line (central temperature and density conditions at which the energy release from the  carbon fusion reactions equals the neutrino emissivity). This time is shorter than the  timescale needed by the WD to reach mass-shedding, secular axisymmetric instability and/or inverse $\beta$ decay instability.
\item
If a magnetized sub-Chandrasekhar WD has a mass which is not very close to the non-rotating Chandrasekhar mass, then it evolves slowing down and on a much longer time scales with respect to SCWDs. It gives us the possibility to observe them during their life as active pulsars. Soft gamma-repeaters and anomalous X-ray pulsars appear to be an exciting possibility to confirm such a hypothesis \citep{2012PASJ...64...56M,2013A&A...555A.151B,2013ApJ...772L..24R,2014PASJ...66...14C,2016IJMPD..2541025L}. The massive, highly magnetized WDs produced by WD binary mergers, recently introduced as a new class of low-luminosity gamma-ray bursts \citep{2018JCAP...10..006R,2018arXiv180707905R} would be another interesting case.
\item
A magnetized WD produced in a WD binary merger will be surrounded by a Keplerian disk \citep{2012ApJ...749...25G}. Thus, to compute its evolution we have to consider the star-disk coupling and the angular momentum transfer from the disk to the WD \citep{2013ApJ...772L..24R}. These new ingredients would appear to invalidate our assumption of an isolated WD and the general picture drawn in this work. However, as shown by \citet{2013ApJ...772L..24R}, accretion from the disk occurs in very short time scales and the WD lives most of its evolution in the dipole magnetic braking phase. Thus, we expect the WD lifetime computed in this work to be a good estimate also for those systems. In that case it is important to estimate the lifetime for the final WD mass after the accretion process since, as we have shown in Fig.~\ref{fig:lifetime}, even a small increase in mass shortens drastically the lifetime.
\end{enumerate}

\section*{Acknowledgement}
\medskip
\noindent
The work was supported by the Ministry of Education and Science of the Republic of Kazakhstan, Program IRN: BR05236494.

\end{document}